\documentclass[letter]{aa} 
\usepackage{graphicx}
\usepackage{txfonts}
\usepackage{natbib}
\newcommand{\um}{\,$\mu$m}
\newcommand{\kms}{km\,s$^{-1}$}
\newcommand{\cii}{[C\,{\sc ii}]}
\newcommand{\ci}{[C\,{\sc i}]}
\newcommand{\hii}{H\,{\sc ii}}

\newcommand{\cc}{cm$^{-3}$}
\bibpunct{(}{)}{;}{a}{}{,}
\begin{document}
   \title{Dynamics and PDR properties in IC1396A}

   \author{Yoko Okada \inst{1}
          \and
          Rolf G\"{u}sten \inst{2}
          \and
          Miguel Angel Requena-Torres \inst{2}
          \and
          Markus R\"{o}llig \inst{1}
          \and
          Paul Hartogh\inst{3}
          \and
          Heinz-Wilhelm H\"{u}bers \inst{4,5}
          \and
          Thomas Klein \inst{2}
          \and
          Oliver Ricken \inst{1}
          \and
          Robert Simon \inst{1}
          \and
          J\"{u}rgen Stutzki \inst{1}
          }

   \institute{I. Physikalisches Institut der Universit\"{a}t zu K\"{o}ln, Z\"{u}lpicher Stra{\ss}e 77, 50937 K\"{o}ln, Germany
              \email{okada@ph1.uni-koeln.de}
              \and
              Max-Planck-Institut f\"{u}r Radioastronomie, Auf dem H\"{u}gel 69, 53121 Bonn, Germany
              \and
              Max-Planck-Institut f\"{u}r Sonnensystemforschung, Max-Planck-Stra{\ss}e 2, 37191 Katlenburg-Lindau, Germany
              \and
              Deutsches Zentrum f\"{u}r Luft- und Raumfahrt, Institut f\"{u}r Planetenforschung, Rutherfordstra{\ss}e 2, 12489 Berlin, Germany
              \and
              Institut f\"{u}r Optik und Atomare Physik, Technische Universit\"{a}t Berlin, Hardenbergstra{\ss}e 36, 10623 Berlin, Germany
             }

   \date{Received; accepted}

 
  \abstract
   {}
   {We investigate the gas dynamics and the physical properties of photodissociation regions (PDRs) in IC1396A, which is an illuminated bright-rimmed globule with internal structures created by young stellar objects.}
   {Our mapping observations of the \cii\ emission in IC1396A with GREAT onboard SOFIA revealed the detailed velocity structure of this region.  We combined them with observations of the \ci\ ${}^3P_1 - {}^3P_0$ and CO(4-3) emissions to study the dynamics of the different tracers and physical properties of the PDRs.}
   {The \cii\ emission generally matches the IRAC 8\um, which traces the polycyclic aromatic hydrocarbon (PAH) emissions.  The CO(4-3) emission peaks inside the globule, and the \ci\ emission is strong in outer regions, following the 8\um\ emission to some degree, but its peak is different from that of \cii.  The \cii\ emitting gas shows a clear velocity gradient within the globule, which is not significant in the \ci\ and CO(4-3) emission.  Some clumps that are prominent in \cii\ emission appear to be blown away from the rim of the globule.  The observed ratios of \cii/\ci\ and \cii/CO(4-3) are compared to the KOSMA-$\tau$ PDR model, which indicates a density of $10^4$--$10^5$ \cc.}
   {}


   \authorrunning{Y. Okada et al.}

   \maketitle
%

\section{Introduction}
\label{sec:introduction}

IC~1396 is a large \hii\ region in the Cep OB2 association, excited by the O6.5V star HD~206267.  Its border includes many bright-rimmed molecular clouds \citep{Weikard1996}, among which IC~1396A is the closest globule, 3.7~pc away from HD~206267 in the plane of the sky assuming a 750~pc distance to IC~1396 \citep{Matthews1979}.  IC~1396A has a more negative velocity ($\sim -8$~\kms) compared to HD~206267 and several of the other bright-rimmed clouds, indicating that it is located in front of the star and moving toward us \citep{Weikard1996}.  IC~1396A thus provides a simple example of photodissociation regions (PDRs) on a globule, illuminated by a single star from one side.  Within the globule a cavity associated with the young stellar object (YSO) LkH$\alpha$ 349 is seen in IRAC 8\um, $^{12}$CO and $^{13}$CO(1-0) maps \citep{Nakano1989}, and the extinction map obtained from The Two Micron All-Sky Survey, 2MASS, data \citep{Reach2009}.  There are additional cavities associated with YSOs, and \citet{Reach2009} suggested that the globule is being reshaped from the inside, with each protostar residing in a ``compartment'' that it has blown out via its outflow.  Therefore, high spatial resolution, as provided by the German REceiver for Astronomy at Terahertz Frequencies \citep[GREAT\footnote{GREAT is a development by the MPI f\"{u}r Radioastronomie and the KOSMA / Universit\"{a}t zu K\"{o}ln, in cooperation with the MPI f\"{u}r Sonnensystemforschung and the DLR Institut f\"{u}r Planetenforschung};][]{Heyminck2012} onboard the Stratospheric Observatory for Infrared Astronomy \citep[SOFIA;][]{Becklin2009,Young2012} at THz-frequencies is essential for understanding the structure and evolution of the IC~1396A globule.

\section{Observation and data reduction}
\subsection{\cii\ observations with SOFIA/GREAT}

\begin{figure*}
\centering
\includegraphics[width=\textwidth]{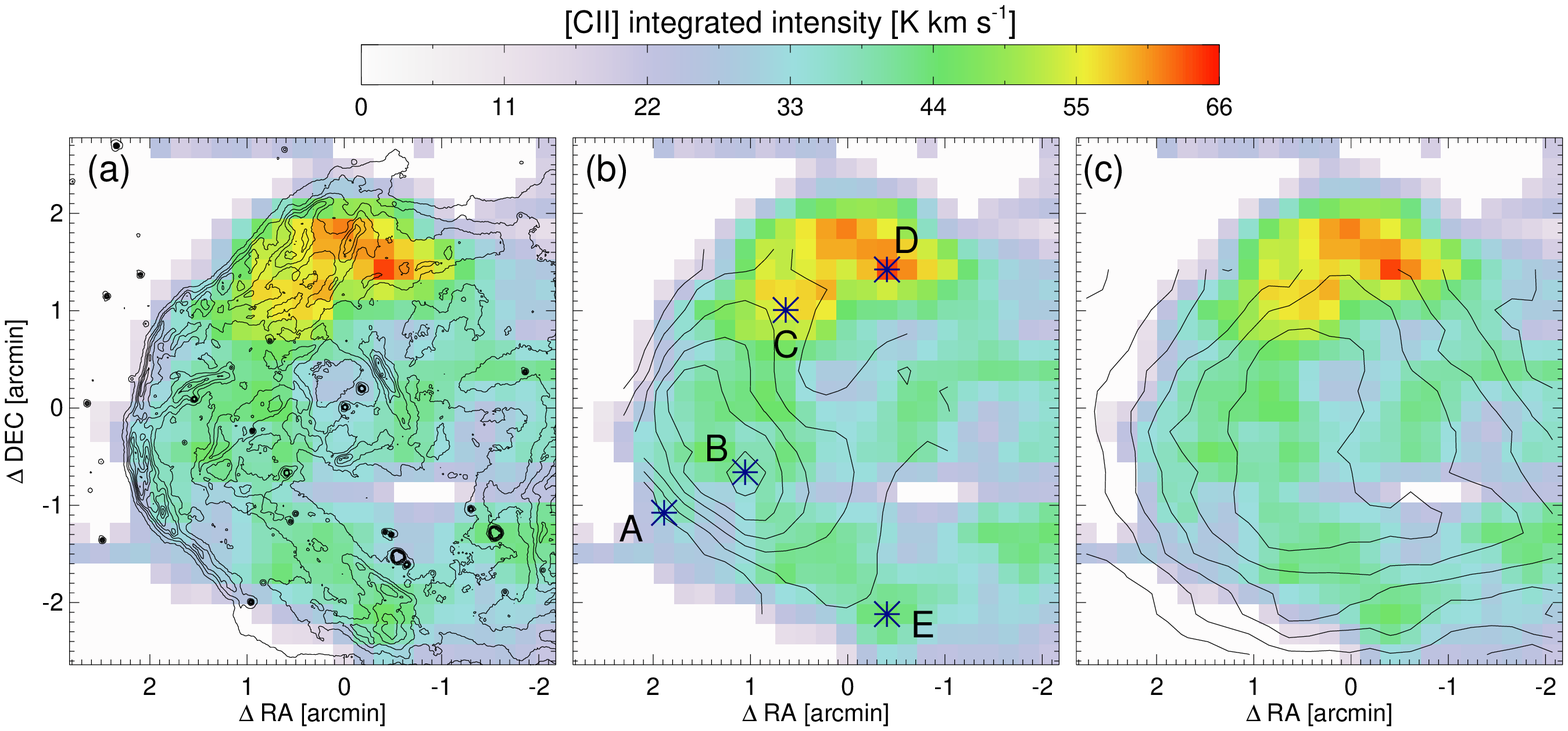}
\caption{Integrated ($-5$ to $-15$~\kms) intensity map of the \cii\ emission (color), overlayed with the contours of (a) IRAC 8\um, (b) velocity-integrated \ci\ ${}^3P_1 - {}^3P_0$ and (c) CO(4-3).  The contour spacing is (a) $10$~MJy\,sr$^{-1}$, (b) 2~K\,\kms, and (c) 10~K\,\kms.  The (0,0)-position is $21^\mathrm{h}36^\mathrm{m}50^\mathrm{s}.7$, $57^\circ31^\prime10^{\prime\prime}$ (J2000) at LkH$\alpha$~349.  Five asterisks in (b) mark the positions of the spectra shown in Fig.~\ref{fig:spectra}.  HD~206267, the exciting star, is located at ($\Delta$RA, $\Delta$DEC)$=$($17.1$\arcmin, $-1.8$\arcmin) and illuminates the globule from the east.}
\label{fig:cii_integmap}
\end{figure*}

We mapped the \cii\ emission at 1900.5369~GHz (158\um) with GREAT onboard SOFIA during the basic science flights on 22 and 28 July 2011. On the second flight we observed only the \cii\ emission with the L2 channel, whereas on the first flight the L1 channel was operated in parallel tuned to CO(12-11) at 1381.9951050~GHz. The observations were made in total-power on-the-fly (OTF) mode, with 1 second integration time at each dump, and 8\arcsec step size.  The OFF position is located east and outside of the globule at $21^\mathrm{h}37^\mathrm{m}27^\mathrm{s}.9$, $57^\circ31^\prime10^{\prime\prime}$ (J2000).

Calibration was made by the standard pipeline \citep{Guan2012}.  The forward efficiency is 0.95, and the beam efficiency is 0.54 and 0.51 for L1 and L2, respectively \citep{Heyminck2012}.  Because the emission line is sufficiently narrow, only a linear baseline was fitted between $-30$ and $-20$~\kms, and between $0$ and $20$~\kms.  After excluding data suffering from strong standing waves and discarding spectra with excess noise, the data were spectrally resampled to a $0.5$~\kms\ resolution and spatially to 25\arcsec\ (0.09~pc at the distance of 750~pc) to obtain a better signal-to-noise ratio (S/N).  In the following, we use the data from the XFFTS-backend, because the other backends show fully consistent and hence redundant data \citep[see ][ for details of the spectrometers]{Heyminck2012}.

Fig.~\ref{fig:cii_integmap} shows the integrated line intensity, summed over the velocity range from $-5$ to $-15$~\kms.  The error in the line-integrated intensity, estimated from the rms noise of the baseline varies between 2--6~K\,\kms\ and depends on the number of times that the map point was observed.  CO(12-11) is not detected, and the upper limit is 4--13~K\,\kms\ (see the appendix for details of the derivation).

\subsection{Complementary observations}

Complementary observations of \ci\ ${}^3P_1 - {}^3P_0$-line at 492.1606510~GHz and CO(4-3) at 461.0407682~GHz were performed between October 2000 and April 2001 using the CHAMP array receiver \citep{Guesten1998} at the CSO. The array with $2\times 8$ pixels was operated in frequency-switching mode (throw $\pm 25$~MHz for CO and $\pm 10$~MHz for \ci, rate $1$~Hz).  CO(4-3) data were recorded on-the-fly, the weaker \ci\ in raster mode. The auto-correlator back-end array in its high-resolution mode provided 2048 channels with $0.15$ \kms\ spectral resolution. Main beams were 15\arcsec\ at 464 GHz and 14.5\arcsec\ at 492 GHz, with respective main beam efficiencies of 0.52 and 0.51 \citep[see ][ for more details of the set-up]{Philipp2006}.   Both maps were convolved to 25\arcsec\ angular resolution.  The integrated line intensities were obtained in the same manner as for the \cii\ emission (Fig.~\ref{fig:cii_integmap}b, c).

\section{Results and discussion}

\subsection{Dynamics of the \cii\ emitting gas}

Figure~\ref{fig:cii_integmap}a shows an overlay of the \cii\ emission and the IRAC 8\um\ emission, which traces polycyclic aromatic hydrocarbons (PAHs) in PDRs, showing a good spatial match between the two.  The \cii\ emission also follows the local cavities originating from the YSO outflows mentioned in Sect.~\ref{sec:introduction}.  One exception is the rim at the eastern edge of the globule, where the IRAC 8\um\ has a strong peak while the \cii\ emission is not prominent.  The spatial distributions of \ci\ and CO(4-3) are different from \cii.  CO(4-3) is widely distributed over the globule, having a shallow peak in the middle.  The \ci\ emission is distributed toward the southeast and farther out compared to CO(4-3), following the ridge-like structure from ($\Delta$RA, $\Delta$DEC)$\sim$($1$\arcmin, $0$\arcmin) to ($0$\arcmin, $-2$\arcmin) in the IRAC 8\um\ emission. The \cii\ emission also follows this rim, but it is much stronger toward the northern part of the globule.

\citet{Wootten1983} showed the intensity distribution and the position-velocity diagram of CO(2-1) and $^{13}$CO(2-1) along the longitude cut that intersects LkH$\alpha$~349 (located at offset (0,0)).  The CO(2-1) distribution is similar to that of CO(4-3), and $^{13}$CO(2-1) is similar to \ci\ and \cii\ in the sense that the intensity drops around LkH$\alpha$~349.  On the other hand, \citet{Nakano1989} showed the cavity near LkH$\alpha$~349 both in CO(1-0) and $^{13}$CO(1-0) although it appears as a much steeper hole in the latter.  These differences are compatible with being caused by opacity: $^{12}$CO is more optically thick than $^{13}$CO, and $^{12}$CO(2-1) and (4-3) emissions have a higher optical depth than $^{12}$CO(1-0) in LTE gas with $T\gtrsim 10$~K, although detailed radiative transfer modeling is needed to quantify it.

\begin{figure}
\centering
\includegraphics[width=0.49\textwidth]{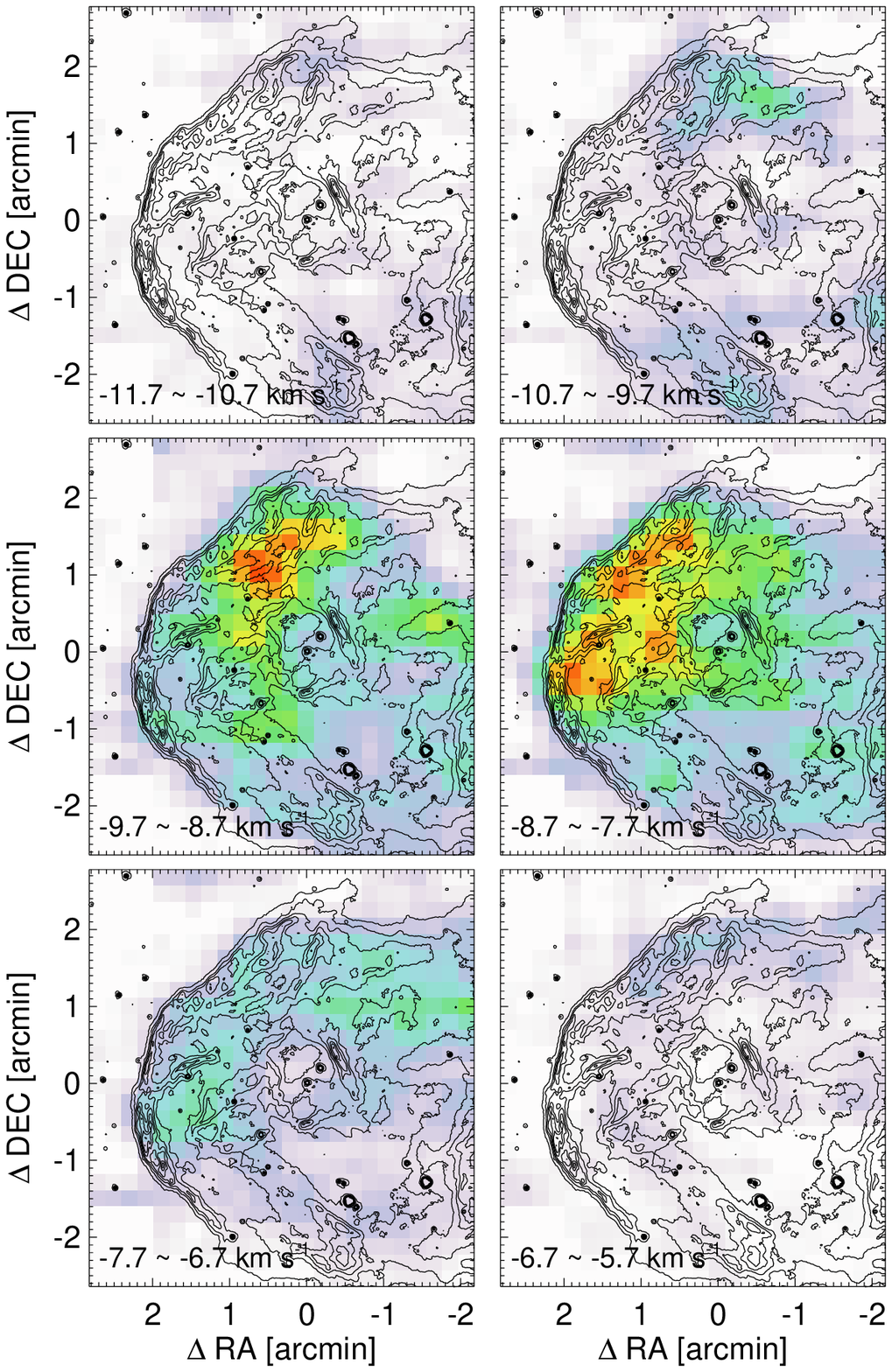}
\caption{Channel maps of the \cii\ emission overlayed with the contours of IRAC 8\um\ emission.}
\label{fig:channelmap}
\end{figure}

\begin{figure}
\centering
\includegraphics[width=0.4\textwidth]{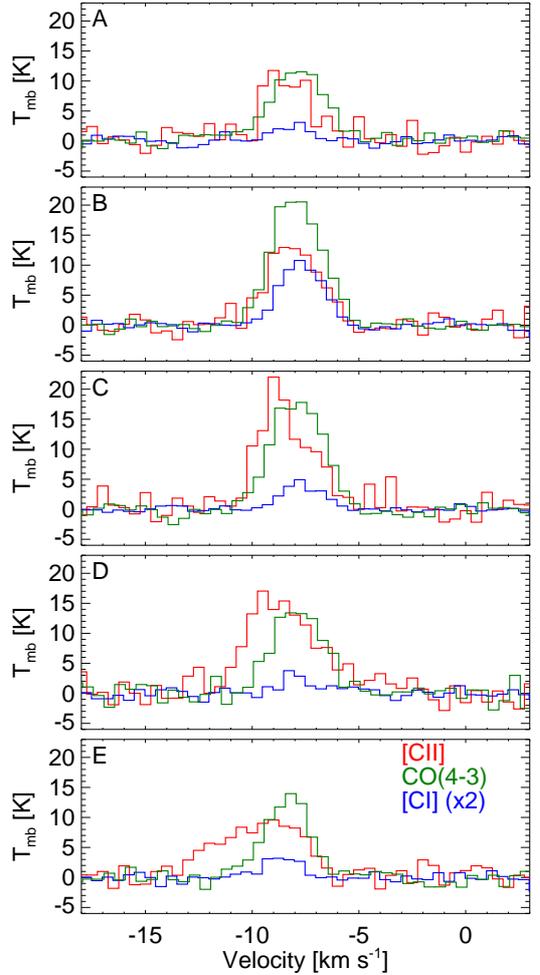}
\caption{Spectra of \cii, CO(4-3) and \ci\ at the five positions marked in Fig.~\ref{fig:cii_integmap}.}
\label{fig:spectra}
\end{figure}

The channel map of \cii\ (Fig.~\ref{fig:channelmap}) shows that the gas near the rim of the globule is redshifted; going west and toward the center of the globule the velocity becomes more negative.  With the globule in front of the exciting star and moving toward us as mentioned in Sect.~\ref{sec:introduction}, this velocity gradient implies that the western part, which is farther away from the illuminated rim, moves faster away from the exciting source.  The upper right panel of Fig.~\ref{fig:channelmap} shows two \cii\ components with significantly higher negative velocity (see also the spectra at positions D and E in Fig.~\ref{fig:spectra}).  In the southern component (position E) the local outflows from the YSOs may also affect the velocity field because there are several YSOs in the void-like structure from ($\Delta$RA, $\Delta$DEC)$\sim$($1.5$\arcmin, $0$\arcmin) to ($0.5$\arcmin, $-2$\arcmin) \citep{Reach2009}, and the IRAC 8\um\ emission shows a cometary shape toward the southwest.

On the other hand, the \ci\ and CO(4-3) emitting gas does not show a significant velocity gradient.  Figure~\ref{fig:spectra} shows the line profiles at the five positions marked in Fig.~\ref{fig:cii_integmap}b.  CO(4-3) has a slight blueshift at position E, which is also indicated by \citet{Wootten1983} in the CO(2-1) position-velocity diagram along the latitude cut.  However, there is a clear difference between the line profiles of \cii\ and CO(4-3).  Considering the spatial distribution discussed above, it seems that the gas emitting \cii\ is blown away from the rim.  The numerical simulations of the evolution of cometary globules by \citet{Lefloch1994} suggest that such a cometary regime lasts $\sim$ 90\% of the cloud's lifetime, and numerous small clumps that are ejected and accelerated along the globule appear as ``fuzz'' on the blue side of Lefloch \& Lazareff's model position-velocity diagram.  The observed structures match this cometary regime.

\subsection{Comparison with PDR models}

\begin{table*}
\caption{PDR model results.  Values in parentheses show the ranges defined by the uncertainties.}
\label{table:modelresults}
\centering
\begin{tabular}{|c|ccc|cc|}
\hline
position & \multicolumn{3}{c|}{Model 1 (fixed $\chi$)} & \multicolumn{2}{c|}{Model 2 (fixed $\log(m)=-1$)}\\
\cline{2-6}
 & $\log(\chi)$ & $\log(n)$ & $\log(m)$ & $\log(n)$ & $\log(\chi)$\\
\hline
A & $2.0$ & $5.0$ ($4.5$ -- $5.4$) & $-2.3$ ($-3.0$ -- $+ 0.8$) & $5.2$ ($4.2$ -- $6.1$) & $ 2.5$ ($ 1.4$ -- $ 3.1$) \\
B & $1.9$ & \ \ -- \ \  ($4.4$ -- $5.1$) & \ \ -- \ \   ($ 0.0$ -- $+ 3.0$) & $4.1$ ($3.8$ -- $5.0$) & $ 1.0$ ($ 0.6$ -- $ 1.8$) \\
C & $1.9$ & $4.7$ ($4.3$ -- $5.3$) & $-0.8$ ($-3.0$ -- $+ 1.7$) & $4.7$ ($4.0$ -- $5.7$) & $ 1.9$ ($ 1.2$ -- $ 2.7$) \\
D & $1.8$ & $4.9$ ($4.3$ -- $5.1$) & $-2.9$ ($-3.0$ -- $+ 0.2$) & $5.0$ ($4.0$ -- $5.9$) & $ 2.5$ ($ 1.3$ -- $ 3.1$) \\
E & $1.8$ & $4.9$ ($4.4$ -- $5.2$) & $-2.0$ ($-3.0$ -- $+ 0.8$) & $5.0$ ($4.0$ -- $5.9$) & $ 2.2$ ($ 1.2$ -- $ 2.8$) \\
\hline
\end{tabular}
\end{table*}

We compared the observed integrated intensity ratios between \cii, \ci\ ${}^3P_1 - {}^3P_0$, and CO(4-3) with the KOSMA-$\tau$ PDR model \citep{Roellig2006} at the five positions shown in Fig.~\ref{fig:spectra}.  To exclude the contribution of the velocity components in \cii\ that do not match with the CO and \ci\ emission, we fitted the \cii\ emission line with the CO(4-3) line profile at each position and adopted the integrated intensity of this fit result.  At position E, the intensity obtained this way is 57\% of the total integral of the \cii.  We used a single-clump model, which has as free parameters the mean gas density ($n$), the far-ultraviolet (FUV; $h\nu =$ 6--13.6~eV) flux ($\chi$) in units of the Draine field ($2.7\times 10^6$~W\,m$^{-2}$), and the mass of the single clump ($m$) in units of the solar mass ($M_\odot$).  We used two simplified model approaches because with three observed lines, there are only two independent ratios but three unknowns.  In Model~1, we fixed $\chi$ at the value estimated from the luminosity of the exciting star HD~206267, $10^{5.23}L_\odot$, as an O6.5V star, and the distance to this star from each observed position, assuming that 50\% of the photons have an energy within 6~eV $< h\nu <$ 13.6~eV.  The calculated value of $\chi$ is listed in Table~\ref{table:modelresults}.  Then we determined $n$ and $m$ by fitting the intensity ratios of \cii/\ci\ and \cii/CO(4-3), as listed in Table~\ref{table:modelresults}.  To be conservative, we estimated the error by assuming that each line intensity has a systematic uncertainty of 30\%.  At position B, no combination of $n$ and $m$ can explain the observed ratio.  $m$ is not well constrained because these ratios are insensitive to the mass; the layers that emit \cii, \ci\ ${}^3P_1 - {}^3P_0$, and CO(4-3) emissions are not the central core but outer layers.  With $\chi=10^2$ and $n=10^5$\cc, $A_V$ at the center is $\sim 4$ even for a clump with $m=10^{-3}M_\odot$.  Increasing $m$ corresponds to adding layers at deeper $A_v$, which does not significantly change the ratio between the \cii, \ci\ ${}^3P_1 - {}^3P_0$, and CO(4-3) intensities.  Therefore, we fixed $\log(m)$ in Model~2 to a value of $-1$, i.e.\ $0.1M_\odot$, and fitted $n$ and $\chi$.  A clump with $m=0.1M_\odot$ and $n=10^5$\cc\ has a radius of $\sim 0.02$~pc, which is $\sim 5\arcsec$ at the distance of 750~pc, small compared to the beam size.  The obtained $n$ from both models are consistent within the errors, giving $10^4$--$10^5$\cc\ at position B and $\sim 10^5$\cc\ for the other positions.  The fitted $\chi$ in Model~2 is consistent with the values estimated from the luminosity of the star for Model~1 except for position B.  At position B, the very low \cii/\ci\ ratio compared to the other positions is explained by a low UV field in Model~2.  It implies that the UV radiation is shielded at the rim of the globule.

\section{Summary}

We presented mapping observations of the \cii\ emission at 1900.5369~GHz (158\um) with GREAT onboard SOFIA in IC1396A, which is an illuminated globule with internal structures created by embedded YSOs.  The \cii\ emission closely follows the IRAC 8\um\ emission, which traces the PAH emission.  Together with complementary \ci\ ${}^3P_1 - {}^3P_0$ and CO(4-3) observations, we investigated the spatial distributions of the velocity-resolved emission.  \cii\ shows significant velocity changes within the globule, while this is not the case for the \ci\ and CO(4-3) emission.  The spatial distribution is also different: CO(4-3) has a shallow peak at the center of the globule, the \cii\ and \ci\ emission is strong at outer region of the globule, but their peaks show different locations. The spatial structure and velocity distribution is consistent with a scenario in which spatially unresolved clumps, which emit mainly \cii, are blown away along the globule and are accelerated toward us.  A PDR model analysis of the observed intensity ratios indicates densities of $\sim 10^5$ \cc\ and a UV flux consistent with excitation by the external star except for one position, where the UV is indicated to be low because of the shielding at the rim of the globule.  IC1396 is selected as one of the key regions to be observed by the GREAT consortium because it provides many bright-rimmed globules with simple exciting sources, and observations in wider regions are expected in future flights.


\begin{acknowledgements}
This work is based in part on observations made with the NASA/DLR Stratospheric Observatory for Infrared Astronomy. SOFIA Science Mission Operations are conducted jointly by the Universities Space Research Association, Inc., under NASA contract NAS2-97001, and the Deutsches SOFIA Institut under DLR contract 50 OK 0901.  We thank the SOFIA engineering and operations teams whose support has been essential for the GREAT accomplishments during basic science flights, and the DSI telescope engineering team.
\end{acknowledgements}

\bibliographystyle{aa}

\begin{appendix}
\section{Estimate of uncertainties}
We estimate the uncertainty or the upper limit of the integrated line intensity from the rms noise ($\sigma$) of the baseline as follows.  $\sigma=1.1$--$2.9$~K and $0.6$--$2.1$~K in the $25\arcsec$ resolution map for \cii\ and CO(12-11), respectively.  At each map position, we create an artificial spectrum by generating normally distributed (Gaussian) random numbers with a standard deviation of $\sigma$.  Then we integrate this spectrum over the velocity range from $-5$ to $-15$~\kms, which is the same range to obtain the line integrated intensity, and obtain an intensity $I$ (K\,\kms).  After repeating this trial 20000 times, we obtain a clear Gaussian distribution of $I$, and adopt its standard deviation ($\sigma_2$) as an uncertainty of the integrated line intensity at this map position.  If $3\sigma_2$ is higher than the integrated line intensity, we consider it a non-detection, and adopt $3\sigma_2$ as an upper limit.
\end{appendix}

\end{document}